\documentclass[prd,twocolumn,showpacs,superscriptaddress]{revtex4-1}

\usepackage{amssymb,amsmath,epsfig}

\def\eq#1{Eq.~(\ref{#1})}
\def\fig#1{Fig.~\ref{#1}}
\def\sec#1{Sec.~\ref{#1}}

\begin{document}

\title{Domain structures in quantum graphity}

\author{James Q. Quach} \email{quach.james@gmail.com}
\affiliation{School of Physics, The University of Melbourne, Victoria 3010, Australia}

\author{Chun-Hsu Su}
\affiliation{Department of Infrastructure Engineering, The University of Melbourne, Victoria 3010, Australia}

\author{Andrew M. Martin}
\affiliation{School of Physics, The University of Melbourne, Victoria 3010, Australia}

\author{Andrew D. Greentree}
\affiliation{School of Physics, The University of Melbourne, Victoria 3010, Australia}
\affiliation{Applied Physics, School of Applied Sciences, RMIT University, Victoria 3001, Australia }

\begin{abstract}
Quantum graphity offers the intriguing notion that space emerges in the low energy states of the spatial degrees of freedom of a dynamical lattice. Here we investigate metastable domain structures which are likely to exist in the low energy phase of lattice evolution. Through an annealing process we explore the formation of metastable defects at domain boundaries and the effects of domain structures on the propagation of bosons. We show that these structures should have observable background independent consequences including scattering, double imaging, and gravitational lensing-like effects.
\end{abstract}


\maketitle

\section{Introduction}
\label{sec:Introduction}

A single theory that reconciles quantum mechanics with general relativity would revolutionise our fundamental understanding of reality. Such an endeavour in the search for a quantum theory of gravity has traditionally been approached either through the quantization of general relativity or extensions of quantum field theory. For example loop quantum gravity~\cite{rovelli2004quantum,*thiemann2001introduction,*ashtekar2004background}, spin foam models~\cite{bianchi2006graviton}, causal dynamic triangulation (CDT)~\cite{ambjorn2006quantum}, and group field theory~\cite{oriti2006group,oriti2007group} fall into the first category and collectively the string theories fall into the second~\cite{green1988superstring,*weinberg1996quantum,*polchinski1998string,*ibanez2000second}. Condensed matter physics provides a third conceptual framework, where gravity is viewed as an emergent phenomenon. Motivated by techniques and concepts native to the study of many-body interactions in condensed matter physics, this perspective has recently stimulated investigations into gravity analogues in condensed matter systems~\cite{unruh1981experimental,calzetta2003bose,fedichev2003gibbons,visser2007analogue,volovik2009universe,menicucci2010simulating}, emergent graviton excitations in qubit models~\cite{gu2009emergence}, and quantum graphity (CDT and matrix models~\cite{banks1999tasi,horava2005stability} may also be viewed from this perspective).

Motivated by the removal of presumptions of the nature of spacetime, quantum graphity (QG) was proposed as a model in which ideas such as continuity, dimensionality, and macro locality of the spacetime manifold are emergent phenomena. Previous works have studied how locality emerges in the model~\cite{konopka2008quantum}, the role matter plays in the emergence of extended geometries~\cite{konopka2008statistical}, the entanglement of matter with spatial degrees of freedom~\cite{hamma2010quantum}, Ising mappings to study low temperature properties~\cite{caravelli2011properties}, and the entrapment of matter in regions of high connectivity~\cite{caravelli2011trapped}. See Ref.~\cite{hamma2011background} for a review. 

In QG it is purposed that a low dimensional regular graph, representing flat local space, emerged from an early universe, represented by a complete graph. As every vertex is connected to every other, there is no notion of subsystems and hence no notion of locality, which Konopka~\emph{et~al.}~\cite{konopka2008quantum} interprets as representing a state with no space. The evolution from the complete graph of the early QG universe to a lower energy state of the spatial degrees of freedom is reached by the destruction (and creation) of edges. In the open (non-unitary) QG model~\cite{konopka2008statistical}, it is assumed that the lattice is connected to an external heat bath through which edges are exchanged. Under particular parametrical constraints it was shown that the model can give rise to an \emph{hexagonal} or \emph{honeycomb} lattice as a stable local minimum~\cite{konopka2008quantum}. This graph has the desirable properties that the spatial degrees of freedom are local and low dimensional. Here we explore metastable defects to this local minimum state and investigate the possible observable effects on the propagation of bosons.

The study of defects is important, as in most realistic or non-idealised systems, ranging from ferromagnet to field theoretic particle and cosmological models, metastable defects in some form exist. In many field theoretic early universe models for example, topological defects are an unavoidable causal process, and have been proposed to play an important role in cosmic evolution~\cite{vilenkin2000cosmic}. In particular, cosmic strings have been proposed as the seeds for large-scale structure formation such as galaxies, offering a theoretical alternative to inflation~\cite{zel1980pis,vilenkin1981cosmological}. However observational data indicating that cosmic microwave background anisotropies significantly differ from the anisotropies that are predicted to be the result of topological defects, may relegate defects to a subsidiary role in cosmic structure formations~\cite{vilenkin2000cosmic}. In large extra dimensional models, topological defects provide a mechanism by which 3+1 dimensional branes exist in a higher dimensional bulk~\cite{akama1983lect,rubakov1983we,arkani1998hierarchy}.

Topological defects, of which the above are examples, occur as the result of the spontaneous choice of an order parameter value that breaks a symmetry of the system. Defects in QG on the other hand, of the type that are disruptions to the ordered crystal structure, are more akin to crystallographic defects. In this context, QG can be treated analogously to crystallographic models of condensed matter physics. Of course there are important differences. In conventional crystallographic models, the structures' internal interconnectivity is determined by the spatial arrangement of the atoms, which changes with the motion of the atoms. QG has an almost antithetical perspective: It is the interconnectivity of spatial points that determines spatial separation. The dynamics of the QG lattice is not viewed as the movement of these spatial points, but the change in the connections that relate these points. Moreover these interconnectivities in QG are quantum degrees of freedom as they can be in a superposition of \emph{on} or \emph{off} states. Notwithstanding these differences, one can draw similarities between the evolution of the QG universe and the crystallization of solids.

The ordering of unit cells into periodic structures (crystallization) is a well-known phenomenon. Although occurring in a wide variety of systems (e.g. freezing water, cooling magma), the process is always essentially the same: Random atomic distributions nucleate ordered structures below some critical temperature. And if this low temperature is maintained, these seeds will continue to grow and coalesce into larger crystal structures. The final atomic arrangement will be dependent on the initial configuration prior to cooling, rate of cooling, and random effects of thermal fluctuations. Typically a slower cooling rate will produce larger grains. Due to imperfections and thermal fluctuations however, for most materials it is unlikely that the process will end in a structure without defects (although under very controlled environments single crystals, such as monocrystalline silicon, can be manufactured). Instead the material will most probably settle into a metastable (local minimum) state with many grains, resistant to thermal or mechanical perturbation. 

One may view the complete graph of the early universe analogously to the high temperature diffused state of heated crystalline structures, and subsequently see in the QG model similar recrystallization qualities in the cooling phase. Specifically, fluctuations may seed the nucleation of space-like separated local energy minimum regions which are subgraphs of the ground state. The growth of these regions will give rise to domains (grains). As with the cooling of heated solids, it is not inevitable that the domains will coalesce to a global ground state. Analogous to conventional crystallization processes, the ripening (growth) of the different domains will see a competition for energetically favourable local configurations, resulting in a granular structure of space. 

Defects in the space manifold will affect particle propagation. On cosmological scales the classical treatment of the influence of gravity on matter has been sufficient in yielding accurate predictions: from the deflection of particles near massive objects to gravitational redshifts. The fact that on large scales classical treatments accurately account for observation provides an important avenue to test QG. In particular, here we explore the possible observational consequences of domain structures and defects in a classical lattice for the propagation of particles.

In \sec{sec:Model Setup} we briefly review and define the scope of the QG model to be considered. In \sec{sec:Antiphase boundary defects} we investigate the formation of metastable domain structures. Through an annealing process we show that defects that arise from an unstable interface of antiphasing domains are metastable. In \sec{sec:grain boundarys} we examine the effects these structures have on boson propagation. We further explore other metastable domain structures that reveal interesting effects, leading to possible observable consequences.

\section{Model Setup}
\label{sec:Model Setup}

\begin{figure}%
	\centering
	\includegraphics[width=0.2\columnwidth]{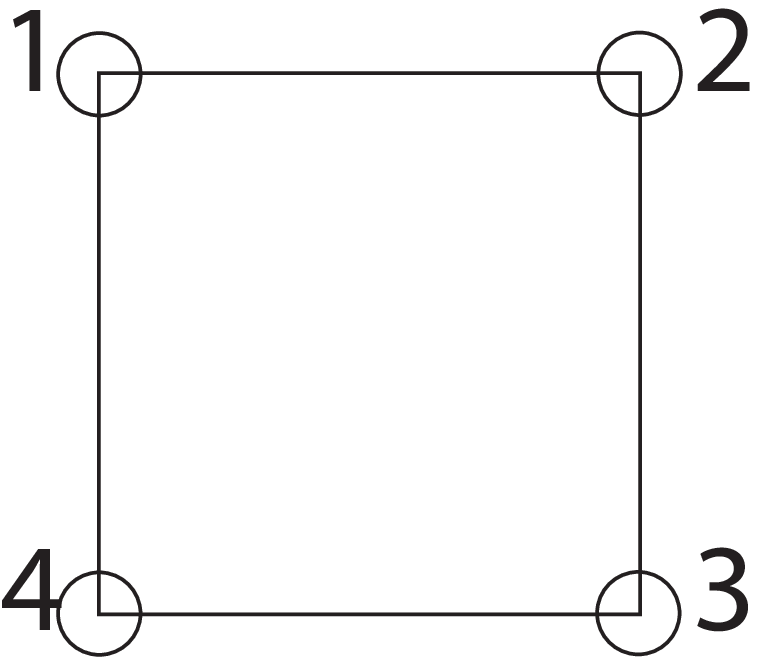}%
	\caption{A graph representation of the state vector $|l_{12}l_{13}l_{14}l_{23}l_{24}l_{34}\rangle\ = |101101\rangle$ which has for 4 vertices and edges.}%
\label{fig:square_single}%
\end{figure}

The Hilbert space of the spatial degrees of freedom in the QG lattice is the tensor product of the individual state vector edge $|l_{rs}\rangle \in \{|0\rangle, |1\rangle\}_{r,s}$.  $|0\rangle_{r,s}$ indicates no edge and $|1\rangle_{r,s}$ an edge between vertices $r$ and $s$.  The set spanned by $\{|0\rangle,|1\rangle\}$ constitutes an orthonormal basis. We define creation and annihilation operators on this space as $b_{rs}^\dagger \equiv |1\rangle\langle 0|_{r,s}$ and $b_{rs} \equiv |0\rangle\langle 1|_{r,s}$ respectively. \fig{fig:square_single} provides an example of a graph representation of the state vector  $|l_{12}l_{13}l_{14}l_{23}l_{24}l_{34}\rangle\ = |l\rangle_{12}|l\rangle_{13}|l\rangle_{14}|l\rangle_{23}|l\rangle_{24}|l\rangle_{34}\rangle\ =|101101\rangle$ in this space. Note only the interconnectivity of the graph is encoded in the state vector. The fact that we have embedded it in a two-dimensional plane and represented it as a square is only an illustrative choice.

Matter degrees of freedom exist on the vertices of the graph. We consider the case of bosons. The Fock space is spanned by their individual vertex Fock states $|n_r\rangle$. Acting on this space are the standard bosonic creation and annihilation operators, $a_r^\dagger$ and $a_r$. The total Hilbert space consisting of the spatial and matter degrees of freedom is spanned by $\{\prod_{r < s}{|l_{rs}\rangle} \otimes \prod_{r'}{|n_{r'}\rangle}\}$. States of the system are represented as a tensor product of the spatial $|\psi\rangle=\sum_{l} c_{l}\prod_{r < s}|l_{rs}\rangle$ and matter $|\phi\rangle=\sum_nc_{n} \prod_{r}|n_{r}\rangle$ degrees of freedom: $|\psi\rangle\otimes|\phi\rangle$. 

The edge number operator, which counts the number of edges at a vertex is defined as $m_{r} \equiv \sum_s{b_{rs}^\dagger b_{rs}}$ and the particle number operator is defined in the standard way, $n_r \equiv a_r^\dagger a_r$. Powers of the edge number operators are also defined as
\begin{equation}
	m_{rs}^{(L)}\equiv\sum_{q_1,...,q_{L-1}}{m_{rq_1}m_{q_2q_3}...m_{q_{L-1}s}}~.
\label{eq:m_power}
\end{equation}
Defined this way, $m_{rs}^{(L)}$ gives the number of paths between vertices $r$ and $s$ of length $L$.

In principle there are  many forms the Hamiltonian $H$, which associates energies to the lattice, can take. However the choice of $H$ is greatly constrained by the QG proposition that locality is an emergent phenomenon. Konopka~\textit{et~al.}~\cite{konopka2008quantum} showed that a Hamiltonian consisting of \emph{valence} term, $H_\text{val}$, that energetically favors vertices with a parameterized number of edges and a $\emph{loop}$ term, $H_\text{loop}$, that discriminately assigns energy to the number of closed loops, can give rise to a stable local minimum state that exhibits locality. In particular, the valence term is given by
\begin{equation}
	H_\text{val} = g_\text{V}\sum_r e^{p(v_0 - m_r)^2}~,
\label{eq:H_V}
\end{equation}
where $p$ is a real positive parameter that penalizes vertex valences (i.e. number of attached edges) that are away from $v_0$, and $g_\text{V}$ is assumed to be positive. The loop term is given by
\begin{equation}	
	H_\text{loop} = -g_\text{P}\sum_{r,L}\delta_{rs}e^{Rm_{rs}}~,
\label{eq:H_L}
\end{equation}
where the coupling parameter $g_\text{P}$ is positive. $e^{Rm_{rs}}\equiv\sum_{L=0}^\infty \frac{R^L}{L!}m_{rs}^{(L)}$ sums all weighted pathways between vertex $r$ and $s$; together with $\delta_{rs}$, only closed pathways (or loops) are counted. The role of parameter $R$ is to influence the loop length $L$ that corresponds to the peak of the effective coupling parameter, $g_\text{P}^\text{eff}=g_\text{P} R^L/L!$. Specifically, $g_\text{P}^\text{eff}$ increases with $L$ for small $L$, but quickly decreases for large $L$; $R$ determines this turnover point. It is of note that as the effective coupling rapidly decreases with large $L$, the contribution from $H_\text{loop}$ can be approximated by truncating \eq{eq:H_L} at some maximal loop length; this is important as the loop counting process is computationally intensive.

\begin{figure}%
	\includegraphics[width=\columnwidth]{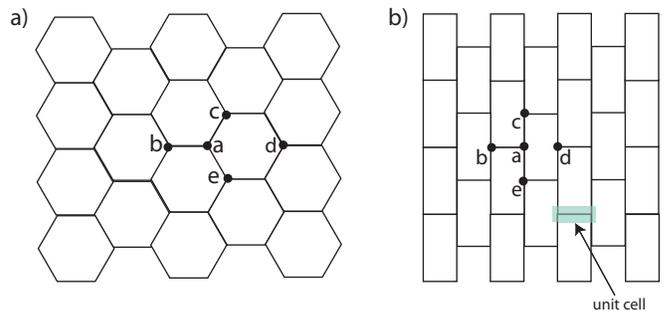}%
	\caption{Two isomorphically equivalent representations of a local minimum graph state for $v_0=3$, $R\ge7.1$: a) Honeycomb representation $[a=(0,0), b=(-1,0), c=(0.5,0.87), d=(2,0), e=(0.5,-0.87)]$ and b) brick representation $[a=(0,0), b=(-1,0), c=(0,1), d=(1,0), e=(0,-1)$. $a,b,c,d,e$ are examples of the respective graphs' vertex labelling schemes.}
\label{fig:honeycomb_brick}
\end{figure}

For the parameter choices $v_0=3$ and $R\ge7.1$ in the regime $g_\text{V} \gg g_\text{P}$, $H_\text{val} + H_\text{loop}$ produces a local minimum honeycomb graph. \fig{fig:honeycomb_brick}a) shows a representation of the honeycomb and \fig{fig:honeycomb_brick}b) an isomorphically equivalent graph which we will call the \emph{brick} representation. $H_\text{val} + H_\text{loop}$  is by no means the only Hamiltonian that will result in low energy graphs with desirable traits such as locality and translation symmetry; other choices may yield different ground states also with such properties. Without loss of generality, we use $H_\text{val} + H_\text{loop}$ ($v_0=3$,$R\ge7.1$, $g_\text{V} \gg g_\text{P}$) as a toy example in a general framework that can be applied to other Hamiltonians, to gain insight into defects in the QG model.

The evolution from the complete graph of the early QG universe to a lower energy state of the spatial degree of freedom is reached by the destruction and creation of edges. The lattice may be assumed to be in contact with an external heat bath through which edges are exchanged~\cite{konopka2008quantum} or in an alternative unitary model, energy is conserved through the coupling of edges to matter degrees of freedom~\cite{hamma2010quantum}. Specifically in this unitary model, the destruction of an edge is accompanied by the creation of two bound quanta. The quanta hop according to
\begin{equation}
	H_\text{hop} = \kappa\sum_{rs} m_{rs} a_r^\dagger a_s,
\label{eq:H_kappa}
\end{equation}
where $\kappa$ is proportional to the hopping frequency and $m_{rs}$ restricts hopping to nearest neighbours. Note that in previous work~\cite{hamma2010quantum}, $t$ is used to denote this coupling parameter. We use $\kappa$ instead, as $t$ will be used to represent time. The bound quanta are then subsequently destroyed creating an edge to form a different lattice topology. 

\begin{figure}%
\includegraphics[width=0.4\columnwidth]{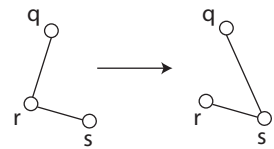}%
\caption{A graphical representation of edge hopping term $H_X = g_X\sum_{qrs} m_{rs}b_{rq}^\dagger b_{qs}$. In this example edge $|l_{rq}\rangle$ is destroyed and edge $|l_{qs}\rangle$ created. This local interaction is only allowed because there is a edge between vertex $r$ and $s$ i.e. $m_{rs}|l_{rs}\rangle=1$.}%
\label{fig:HX}%
\end{figure}

Other possible dynamic interactions include edge exchanges where the valence of vertices are conserved~\cite{konopka2008quantum,konopka2008statistical} and edge hopping. An edge analog of $H_\text{hop}$, edge hopping is a local propagation of edges. We introduce it as
\begin{equation}
	H_X = g_X\sum_{qrs} m_{rs}b_{rq}^\dagger b_{qs}~,
\label{eq:H_X}
\end{equation}
where $g_X$ is a coupling parameter. \fig{fig:HX} illustrates this interaction. In the context of the Hamma~\emph{et~al.}~\cite{hamma2010quantum} model where every edge is assigned some constant energy $U$, every matter quantum assigned energy $\mu$, and the edge-matter coupling constant is $k$, $H_X$ can be considered as an effective dynamic lattice term in the limit $k\ll|U-\mu|$. In this large detuning limit where the difference in the lattice and matter energy scales are much greater than their coupling, the dynamics of the matter and lattice degrees of freedom are effectively decoupled.

Here we investigate defects in the brick lattice. We study the large detuning or decoupled limit where matter dynamics ($H_\text{hop}$) are treated independently of the lattice dynamics ($H_X$) i.e. we ignore the effects of matter-lattice coupling. Furthermore, we restrict our investigation to a semi-classical model, in the sense that lattice dynamics are viewed classically through a heuristic process but the matter degree of freedom is studied quantum mechanically.

\section{Metastable domain boundary defects}

\label{sec:Antiphase boundary defects}
In the QG model, the early high energy high temperature universe is represented with a complete graph which may be interpreted as a state without a notion of space. It evolves to a low energy low temperature state that exhibits translational symmetry and locality. The evolution process from the complete graph to this lower energy state however is difficult to calculate as it amounts to a many-body problem for which the simulation time  grows exponentially with the number of edges and vertices. Further compounding the situation is the problem of counting the number of closed loops. As the number of loops grows exponentially with the connectedness of the graph, this problem becomes increasingly difficult nearer to the early universe.

Although a direct analysis of the evolution of the initial complete graph is impractical for graphs large enough to exhibit internal geometry, we can garner insight by comparison with well known crystallographic systems. The random distribution of atoms, say in metal alloys, in the disordered high energy high temperature state means that the system is rotationally symmetric. Correspondingly for a complete graph with $N$ edges, the system exhibits a discrete \emph{$N$-rotational} symmetry~\cite{caravelli2011trapped}, which approaches a continuous rotational symmetry as $N\rightarrow\infty$. 

As the rotational symmetry of the condensed matter system which exists at high temperatures (energy) is broken as the system cools, this also occurs in QG. Specifically, nucleation sees the ordering of atoms which grow into crystal structures that break rotational symmetry; and as the result of thermal fluctuations and other imperfections, this crystallization process may occur independently in space-like separated regions. It is likely that a similar process occurs in the evolution of the complete QG graph to a lower energy state. It is however unclear whether quantum or thermal fluctuations or an interplay of the two will play the dominant role in the nucleation of low energy symmetry breaking in such crystalline structures.

In condensed matter systems, when there is enough energy to overcome the activation barrier, the coalescence of locally ordered regions can allow the formation of larger crystal structures. However, even with a slow cooling rate, defects are likely to form in finite time. A particular type of defect known as an antiphase boundary defect, occurs when the intersection of two domains are \emph{out of phase}. To illustrate this, consider the crystalline structures of alloys which are interpenetrating lattices of the different constituent atoms. For example alloys of two types of atoms, $A$ and $B$, of composition $AB$ (e.g. CuZn) can form crystal grains of a simple cubic Bravais superlattice. This superlattice is composed of two interpenetrating cubic sublattices of $A$ atoms and $B$ atoms. At grain intersections these sublattices may be out of step or phase, in which case an antiphase boundary defect occurs.

\begin{figure}%
\includegraphics[width=.75\columnwidth]{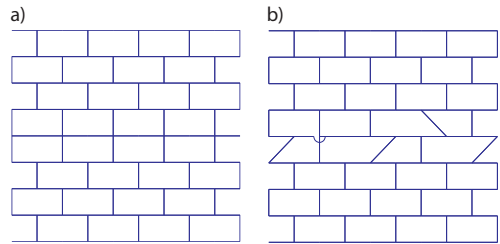}%
\caption{a) Two domains form independently. Where they meet is an unstable antiphase boundary defect. b) After the quenching process the boundary defect is frozen into a metastable amorphous state.}%
\label{fig:annealedWalls}%
\end{figure}

Analogously domains of subgraphs of the ground state may also be \emph{out of phase} at domain boundaries as depicted in \fig{fig:annealedWalls}a). We will also consider examples of when domains differ in orientation in \sec{sec:grain boundarys} and \sec{sec:domain lensing}.  Note that the metal alloy system involves the interleaving of two elements (e.g. Cu and Zn), whereas in the QG brick lattice the interleaving is purely a structural one, i.e. consisting of interleaving repeated structural layers as seen in \fig{fig:annealedWalls}a). For convenience we will refer to the brick graph as the ground state; whether it is actually the ground state or a local minimum does not matter for our purposes. The antiphase boundary represented by \fig{fig:annealedWalls}a) is unstable under lattice interaction. For example, edge hopping described by $H_X$ will see an edge attached to a four edged vertex hop to a neighboring two edged vertex, producing a lower energy state of three edged vertices. In a full quantum mechanical treatment, complexity grows with the product of both the matter and lattice degrees of freedom. To simplify the problem, previous work~\cite{hamma2010quantum} limited the manifold to four vertices, and the matter degrees of freedom were hardcore bosons. With such a small number of vertices, internal geometry is not a feature of the model, and so the lattice energy terms were simplified to only being proportional to the number of edges. Here we are interested in the internal geometry of the lattice and therefore must consider a manifold with many more vertices.  The following section describes a heuristic approach to lattice evolution and formation of metastable defects.

\subsection{Quenching}

We simulate the cooling of the lattice using the \emph{Metropolis} algorithm~\cite{metropolis1953equation}. The \emph{Metropolis} algorithm~\cite{kirkpatrick1983optimization} is an iterative algorithm where in each step the probability $P$ of evolving from state $S$ with energy $E$ to a neighboring state $S'$ with energy $E'$ is
\begin{equation}
	P = \begin{cases} 
		1~, & \mbox{if } E > E' \\ 
		\exp(-\frac{E'-E}{k_BT})~, & \mbox{otherwise } 
	\end{cases}
\end{equation}
where $k_B$ is the Boltzmann constant and $T$ temperature. $S'$ is randomly chosen from the neighborhood of states $\{S'\}$ i.e. the set of states reachable by $S$. Specifically, motivated by the edge hopping term~\eq{eq:H_X}, the dynamics of the lattice are mediated by the following heuristic: at each time step, $h$ random number of edges are  allowed to hop to their nearest neighbors, as depicted in \fig{fig:HX}. $h$ corresponds to a measure of fluctuations. The states $\{S'\}$ reachable by this interaction forms the neighborhood of states.

As the probability distribution of states converges to the Boltzmann distribution for finite temperature on long enough time scales, Konopka \textit{et~al.}~\cite{konopka2008statistical} used the Metropolis algorithm to study the ground state. Here we use the algorithm to study the stability of defects.

\begin{figure}%
\includegraphics[width=\columnwidth]{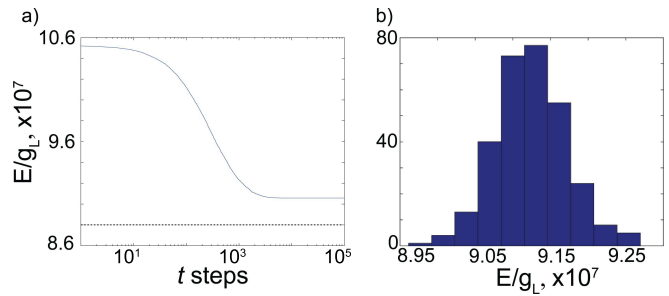}%
\caption{Results of Metropolis algorithm applied to a toroid lattice (major diameter: 100 vertices, minor diameter: 9 vertices) with an unstable antiphase domain boundary [\fig{fig:annealedWalls}a)] at quenching temperature $T=0$. Other parameters values: $v_0=3$, $r=7.2$, $p=1$, $g_\text{V}/g_\text{P}=10^5, L_\text{max}=10,~\text{iterations}=300$.  a) Plot of the lattice energy averaged over 300 samples at each $t$ step. The average energy converges to $9.1\times10^7 E/g_\text{P}$. This is higher than the ground state energy $8.8\times10^7 E/g_\text{P}$ (dotted line), meaning the lattice is more like to settle into a local minimum with defects rather than the ground state. b) The energy probability distribution of the lattice at $t=10^5$, when most of the samples have reached a metastable local minimum. As the lattice has been quenched, the distribution does not follow the Boltzmann distribution for finite $T$, but instead resembles the typical energy distribution of quenched metal alloys.}%
\label{fig:annealedDistributions}%
\end{figure}

We consider the scenario where domains which locally are ground states have formed, and that the temperature has fallen below a critical value such that the system \emph{freezes} into some local minimum. In particular we assume the extreme \emph{quenched} case where the temperature is instantaneously reduced to zero. As the domains formed independently, their interface may be out of phase, for example as illustrated in \fig{fig:annealedWalls}a). This high energy configuration is unstable, and represents our starting state. To remove edge effects, we impose periodic boundary conditions, forming a torus. Along the domain boundary (major diameter) there are 100 vertices, and 9 vertices running along the minor diameter. A small $h$ will mean that the changes to the lattice are local perturbations, leading to the entrapment of metastable defect states in local minima. Applying the Metropolis algorithm with $T=0$, \fig{fig:annealedDistributions}a) shows the decrease in lattice energy, averaged over 300 samples, at each $t$ step of the quenching process. It shows that the average energy converges to $9.1\times10^7 E/g_\text{P}$ and not the ground state energy $8.8\times10^7 E/g_\text{P}$. The parameter values are set to those that give rise to the brick graph as a local minimum, i.e. $v_0=3$, $r=7.2$, $p=1$, $g_\text{V}/g_\text{P}=10^5$. For practical computation times, we have limited the counted maximum loop size, $L_\text{max}=10$, in the above simulations.  We have found no evidence that increasing $L_\text{max}$ significantly affects our results.

\begin{figure}%
\includegraphics[width=1\columnwidth]{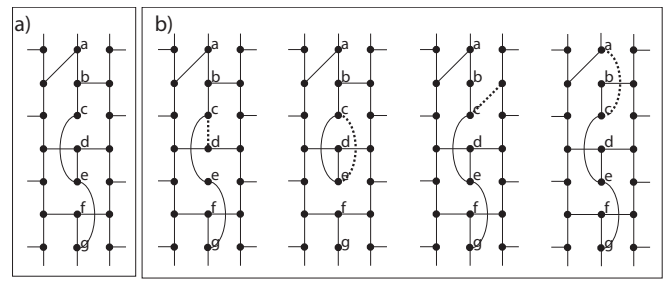}%
\caption{a) An extract of a metastable lattice at the domain boundary after quenching. b) Energetically nearby states accessible from configuration a), by nearest neighbor edge hopping. The dotted lines represent hopping edges. Other neighborhood states (not shown) are of much higher energies as more than one vertices have $v\neq v_0$. Calculation of $H_\text{loop}$ shows that the states in b) are higher energy states then configuration a). Therefore as configuration a) is energetically lower than its set of neighboring states, configuration a) forms a metastable state. }%
\label{fig:evolution}%
\end{figure}

There are many ways the lattice can evolve; moreover, under the extreme case of quenching the lattice will freeze into a state which is not the ground state. For example, \fig{fig:evolution}a) shows a local part of a metastable configuration after quenching with $h=1$. In this local area, the valence of vertex $c$ is $v_c=2$ and all other vertices have valence $v_0=3$. In the $g_\text{V} \gg g_\text{P}$ regime, energetically nearby states are those where an edge has hopped to vertex $c$ so that $v_c = v_0$; all other neighboring states have more than one vertices with $v\neq v_0$ and hence are much higher energy states. There a four energetically nearby states, as shown in \fig{fig:evolution}b); however their $H_\text{loop}$ is smaller than \fig{fig:evolution}a), meaning that they have higher energies. Therefore the lattice freezes in configuration \fig{fig:evolution}a), which is in a much higher energy state than the ground state due to the vertices which do not have valence $v_0$. Note that if $T>0$, then there is a finite probability of evolution to higher energy states; in this case the lattice would eventually evolve to the ground state.

The energy probability distribution of the local minimum states reached by this quenching process is given in \fig{fig:annealedDistributions}b). As expected, it shows that the most probable state is centered around $9.1\times10^7 E/g_\text{P}$. Treating the problem as a statistical ensemble, we do not study the many possible individual metastable configurations. Nevertheless in general, the structures of these metastable states resemble two domains separated by an amorphous boundary defect.  \fig{fig:annealedWalls}b) and \fig{fig:evolution}a) represent extracts of two typical examples. In the local lattice region of configuration \fig{fig:annealedWalls}b), all vertices have valence $v_0$, therefore no further edge hopping is possible in this region, as doing so one would go to a higher energy state. In this local region, \fig{fig:annealedWalls}b) is energetically separated from the ground state by the difference in loop contribution, $H_\text{loop}$. As already discussed above, \fig{fig:evolution}a) is also metastable. Note that in the $g_V\gg g_P$ regime, \fig{fig:evolution}a) is a much higher energy state than \fig{fig:annealedWalls}b). Inspecting the set of local minimum states of the numerical simulation, shows that there are a very low number of isomorphic graphs~\footnote{The Matlab function \emph{graphisomorphism} was used to determine the isomorphism amongst the associated adjacency matrices}.  This indicates that the number of local minima in this energy landscape is large.  As $T=0$ the distribution is not the Boltzmann probability distribution of finite temperatures, but instead resembles the quenched energy distributions of, for example, amorphous metal-metaloid alloys~\cite{weber1985local}. In the Boltzmann probability distribution, lower energy states are more likely to form, with the ground states the most probable, which for the $100\times9$ torus lattice corresponds to a ground state energy of $8.8\times10^7~E/g_\text{P}$. In comparison, in the quenched probability distribution the most likely states are centered around $9.1\times10^7~E/g_\text{P}$. In other words, under the prescribed conditions, stable local minimum defect states are more likely to form than the ground state.

\section{Effects of domain structures on the propagation of bosons}
\label{sec:grain boundarys}
The presence of metastable defects affect the propagation of particles. This is important as the observation of these effects could serve as a possible avenue through which to test QG. In this section we discuss the dispersion of bosons and simulate their propagation in the presence of defects. We study domain structures that give rise to scattering, reflection, refraction, and lensing-like behaviour.

\subsection{Scattering}
\label{sec:Scattering}

In calculating the dispersion relation of bosons on the lattice we are only interested in the hopping term $H_\text{hop}$. We label the vertices of the ground state lattice with $\mathbf{r}_r\equiv (x_r,y_r) \in \mathbb{N}^2$. For a useful mapping between Euclidean space and the lattice, we associate a hop between nearest neighbors with one spatial unit under the Euclidean distance function $d(\mathbf{r}_1,\mathbf{r}_2)=\sqrt{(x_2-x_1)^2+(y_2-y_1)^2}$. This correspondence means that a neighborhood in Euclidean space corresponds to a neighborhood on the lattice. The brick representation as shown in \fig{fig:honeycomb_brick}b) is a \emph{natural} representation under this mapping and vertex labelling scheme. By natural representation we specifically mean that embedding the lattice in Euclidean space, the vertex label corresponds to the Euclidean coordinates. As a comparison, a different labelling scheme, as indicated in \fig{fig:honeycomb_brick}a), sees the honeycomb vertex labels correspond to the coordinates of the embedding Euclidean space. In these examples, Euclidean space is the appropriate embedding space as these lattices represent flat space.  However there are infinitely many possible embeddings and each
will have their corresponding dispersion relation.  Contrasting such embeddings is not the aim of this paper, instead we examine scattering from defects in the brick representation. We however stress that scattering from defects is background independent, although the details of how they are represented are not, and so particle propagation is a useful tool for exploring potential physically testable consequences of the QG model. In fact, detailed measurements of propagating fields may, in the future, allow us to reach some conclusions about how the microscopic structure of spacetime should be viewed.

Forming primitive unit cells as shown in \fig{fig:honeycomb_brick}b), the matrix energy equation can be written as~\cite{datta2005quantum,quach2009band}
\begin{equation}
	\sum_r[H_{rs}]\{\phi_r\} = E\{\phi_s\}~,
\label{eq:submatrix_eq}
\end{equation}
where $[H_{rs}]$ is the $(2 \times 2)$ submatrix of the irreducible representation of $H_\text{hop}$ that relates unit cell $s$ with neighboring unit cell $r\neq s$; $[H_{ss}]$ is the intra-cell interaction. $\{\phi_r\}$ is a $(2 \times 1)$ vector denoting the wavefunction in unit cell $r$. Using Bloch's theorem, this equation can be solved with the ansatz
\begin{equation}
	\{\phi_r\} = \{\phi_0\}e^{i\mathbf{k}\cdot\mathbf{D}_r}~.
\label{eq:bloch_ansatz}
\end{equation}
where $\mathbf{D}_r$ invariantly translates the crystal from some arbitrary unit cell $0$ to unit cell $r$ and $\mathbf{k}\equiv(k_x,k_y)$ is a vector in the space reciprocal to $\mathbf{D}$. 

Substituting \eq{eq:bloch_ansatz} into \eq{eq:submatrix_eq} gives the energy eigenvalue equation
\begin{equation}
	\sum_r[H_{rs}]e^{i\mathbf{k}\cdot(\mathbf{D}_r-\mathbf{D}_s)}\{\phi_0\} = E\{\phi_0\}~,	
\label{eq:submatrix_eq_2}
\end{equation}
which when solved gives the dispersion relation or band structure (see APPENDIX)
\begin{equation}
	[E(\mathbf{k})/\kappa]^2 = 1+4\cos k_x(\cos k_x+\cos k_y)~.
\label{eq:dispersions}
\end{equation}
This dispersion relation is notably anisotropic. However at low energies near $\mathbf{k}=0$ the dispersion relation can be approximated to second-order as, $E(\mathbf{k})\approx\pm\kappa(k_x^2+k_y^2/3-3)$~. This has the quadratic form of a conventional two-dimensional (anisotropic)  free-space dispersion relation. The apparent anisotropy can be approximately eliminated by transforming the lattice (i.e. relabelling the vertices) to the isomorphic honeycomb representation. The dispersion relation in the honeycomb representation is $E^\text{hex} = \pm\kappa\sqrt{1+4\cos \sqrt{3}k_x/2(\cos \sqrt{3}k_x/2+\cos 3k_y/2)}$ which in the low energy limit is approximated as $E^\text{hex}(\mathbf{k})\approx\pm\kappa(3k_x^2/4+3k_y^2/4-3)$, resembling the dispersion relation in conventional two-dimensional (2D) flat space. At higher energies however, higher order terms become significant implying that the dispersion relation in any representation is no longer quadratic, and the anisotropy can not simply be relabelled away. As we live in an isotropic universe (although there is some indication that anisotropies exist on cosmological scales~\cite{bennett148first}), the honeycomb representation may be the appropriate representation in 2D. However to more clearly reveal the effects of the anisotropy of the lattice, we use the brick representation where the anisotropy of the dispersion relation is evident even at low energies.  

To see the effects of the antiphase boundary defects on the propagation of bosons, we simulate a single moded Gaussian pulse centered around $(x_0,y_0)$ with width $(\sigma_x,\sigma_y)$,
\begin{equation}
	|\psi(0)\rangle = N\sum_{x,y}e^{-[(x-x_0)^2/2\sigma_x^2]}e^{-[(y-y_0)^2/2\sigma_y^2]}e^{ik_{x}x}e^{ik_{y}y}|1\rangle_{x,y}~.
\label{eq:gaussian_pulse}
\end{equation}
where $N$ is a normalization factor. Evolution of the system is dictated by the Schr\"{o}dinger equation, $|\psi(t)\rangle = e^{iH_\text{hop}t}|\psi(0)\rangle$. It is important to note here that the evolution of the system is dependent on the lattice interconnectivity and not the representation. 

\begin{figure}%
	\includegraphics[width=\columnwidth]{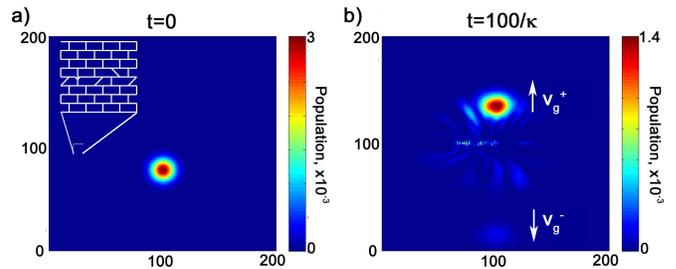}%
	\caption{(Color online) Simulation of a localized boson as it scatters at a metastable amorphous domain boundary defect. a) The boson is initialized at $\mathbf{x}_0=(100,80)$, with $\boldsymbol{\sigma}=(10,10)$ and $\mathbf{k}=(0,-2)$ so that it has $\mathbf{v}_g^\pm/\kappa=\pm(0,1)$. Most of the boson wavepacket propagates as $\mathbf{v}_g^+$. Inset: Zoomed depiction of the amorphous defect line which runs from vertex $(0,100)$ to $(100,100)$. b) Snapshot of the system at $t=100/\kappa$ showing that the amorphous defect has caused the boson to scatter; there is some probability that the boson is also localized on the domain boundary for a finite time.}%
\label{fig:scatterSim}%
\end{figure}

In \fig{fig:scatterSim}a) a Gaussian pulse is initialized at $\mathbf{x}_0=(100,80)$ with $\boldsymbol{\sigma}=(10,10)$ and  $\mathbf{k}=(0,-2)$. The gradient of \eq{eq:dispersions} gives us the group velocity ($\mathbf{v}_g = \nabla_\mathbf{k}E$). As we initialize the boson away from the domain boundaries we can use the dispersion relation, \eq{eq:dispersions}, to estimate the group velocity, $\mathbf{v}_g^\pm/\kappa=\pm(0,1)$. \fig{fig:scatterSim}b) shows that most of the pulse being in the $E^+(0,-2)$ state propagates as $\mathbf{v}_g^+$, and the metastable amorphous defect region has a scattering effect on the boson. Although we have chosen to present the simulation in the brick representation, the manifestation of scattering from domain boundary defects are representation independent consequences of metastability in QG. In other words, choosing a different representation changes only the details of the scattering, and hence scattered bosons may be used by an observer to detect the presence of a domain boundaries. We also note that there is also some localization of the boson along the domain boundary for a finite time.

\subsection{Refraction and reflection}
\label{sec:grain boundarys}

\begin{figure}
\includegraphics[width=.7\columnwidth]{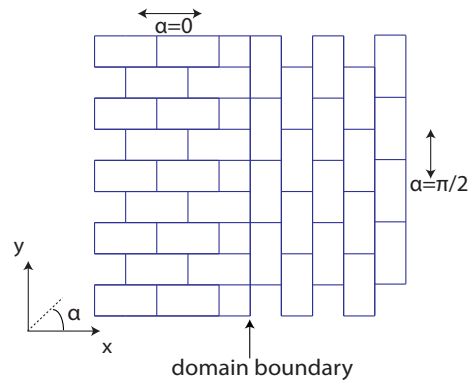}%
\caption{Two domains formed independently with different orientation, labelled as domain-$\pi/2$ and domain-0 . At their intersection is a domain boundary defect.}%
\label{fig:twoDomains}%
\end{figure}

The discrete ordered structure of crystals means that they are not rotationally invariant. In QG, crystallization of space-like separated regions into subgraphs of the ground state, having no preferred orientation, will give rise to dislocations when they meet. \fig{fig:twoDomains} illustrates such an example. The relative orientation of the domains can be represented by a global $\alpha\in[0,\pi)$ rotation of the vertex labels. Within a domain there is short-range order, but this symmetry is broken over longer ranges. We investigate the effects of these types of domain structures on the propagation of bosons.

As an introductory case we consider two domains that are oriented perpendicular to each other as represented in \fig{fig:twoDomains}. This state is metastable under small fluctuations $h$ as there is no nearby state of lower energy i.e. local edge hopping will not produce a lower energy state. We will identify domains by their orientation angle.

\begin{figure}
\includegraphics[width=\columnwidth]{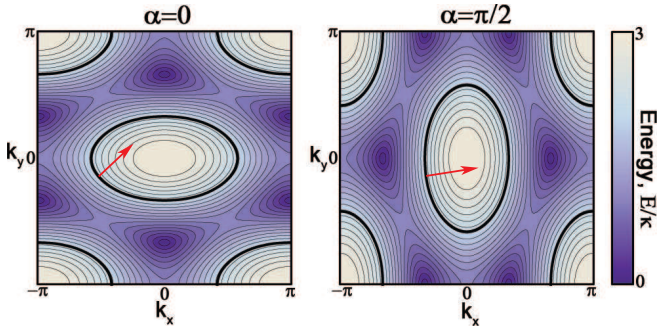}%
\caption{(Color online) Isoenergy contours of the dispersion properties of domain-$0$ (left) and domain-$\pi/2$ (right). A mode with $E/\kappa=2$ (bolded contour lines) and $\mathbf{k}_{0}=(-1.6,-0.5)$ in domain-$0$ has  $\mathbf{v}_g^+/\kappa=(0.88,0.83)$ as indicated by the arrow. Neglecting edge effects, phase matching will mean that this mode will couple to propagating mode $\mathbf{k}_{\pi/2}=(-1.0,-0.5)$ in domain-$\pi/2$, which will have $\mathbf{v}_g^+/\kappa=(1.6,0.3)$ (indicated by arrow). }
\label{fig:refraction_angles}%
\end{figure}

The dispersion relation of the ground state with a generalized rotation of the vertex labels is 
\begin{align}
 &[E({\bf k},\alpha)/\kappa]^2 = 1+4\cos(k_x\sin\alpha-k_y\cos\alpha)\nonumber\\
 	&\quad\times[\cos(k_x\cos\alpha+k_y\sin\alpha)+\cos(k_x\sin\alpha-k_y\cos\alpha)]~.
\label{eq:dispersion_property_general}
\end{align}
\fig{fig:refraction_angles} shows the isoenergy contour plot for $\alpha = 0$ and $\alpha=\pi/2$.

Domain boundaries introduce edge effects, which means \eq{eq:dispersion_property_general} is only valid away from the domain boundary. Nevertheless, as simulations will show, the dispersion relation can be used as a good indicator of the refractive and reflective properties across the domain boundary. 

The difference in the dispersion relation of domains of dissimilar rotation angles will mean that as the boson passes from one domain to another it will experience refraction. Incident and refracted angles are described relative to an interface normal. Using $\tan[\theta(\mathbf{k},\alpha)] = v_{g,y}/v_{g,x}$ and \eq{eq:dispersion_property_general}, the angle of propagation relative to the domain boundary normal is given by
\begin{equation}
	\theta(\mathbf{k},\alpha) = \arctan\Bigl[\frac{c_A s_B s_\alpha + (c_B + 2c_A)s_A c_\alpha}{s_B c_A c_\alpha - (c_B + 2c_A)s_A s_\alpha}\Bigr]~,
\label{eq:theta}
\end{equation}
where $A\equiv k_y c_\alpha - k_x s_\alpha$ and $B\equiv k_x c_\alpha + k_y s_\alpha $. We have used the notation $c_q \equiv \cos q$ and $s_q \equiv \sin q$, $q=A,B,\alpha$.

\begin{figure}%
\includegraphics[width=\columnwidth]{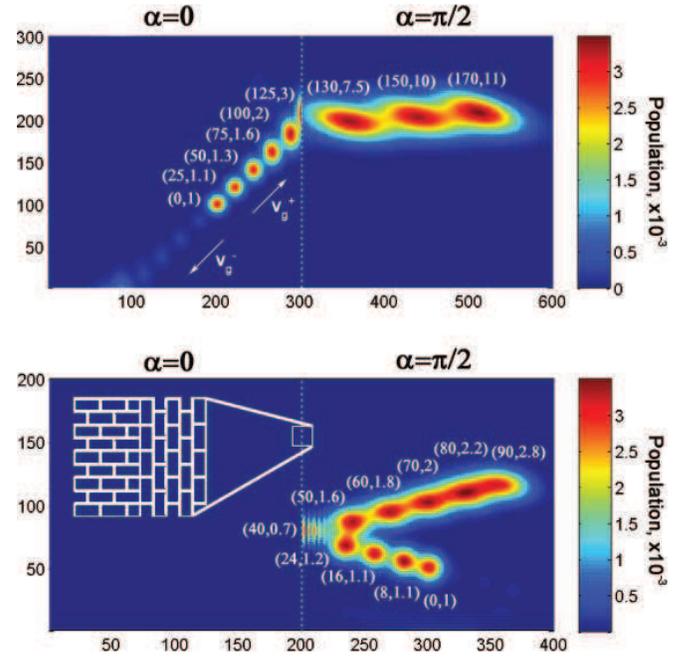}%
\caption{(Color online) Simulation of the propagation of a boson wavepacket over time. Different time instances, $t$, are superimposed; each instance is labeled with
$(\kappa t,M)$. For clearer presentation, populations are multiplied by factor M so that the peak value at each time instant is approximately the same. a) The boson is initialized in domain-0 with $\mathbf{k}_0=(-1.6,-0.5)$. As the boson propagates through the domain boundary it undergoes refraction with angle of refraction $|\theta_R| \approx 8.8^o$ . b) The boson is initialized in domain-$\pi/2$ with $k_{\pi/2}=(0.84,-1.1)$. As this mode can not couple to resonant modes in domain-0 the boson will be completely reflected. Inset: Zoomed depictions of the domain boundary.}%
\label{fig:refraction_simulation}%
\end{figure}

The refractive and reflective properties of the domain boundary are simulated in the propagation of a boson Gaussian wavepacket in a discrete space described by the geometry of \fig{fig:twoDomains}. A boson initiated in domain-0 with $\mathbf{k}_{\alpha=0}=(-1.6,-0.5)$ will have an incident angle given by \eq{eq:theta} as $|\theta_I| = 43^\circ$. As the boson is initialized away from the domain boundary and hence edge effects, this propagation angle is well matched in the simulation as seen in domain-0 of \fig{fig:refraction_simulation}a). This boson will couple to a mode in the domain-$\pi/2$ with the same energy and $k_y$ component. Without edge effects, $\mathbf{k}_{\pi/2}=(-1.0,-0.5)$, giving a refraction angle $|\theta_R| = 8.8^\circ$. Edge effects at the domain boundary will cause deviation from this refraction angle, however this angle can used as a guide as to how light will refract, as shown in \fig{fig:refraction_simulation}a).  

In this toy 2D universe an observer in domain-$\pi/2$ may see two images of an object. This is because for certain incident angles on the domain boundary energy conservation and phase matching can not simultaneously occur and these field components will be reflected to form a second image of the object. \fig{fig:refraction_simulation}b) demonstrates this with $\mathbf{k}_{\pi/2}=(0.84,-1.1)$. The boson initialized at $t=0$ propagates hitting the domain boundary and gets reflected. 

We point out that although the calculated refraction and reflection angles are specific to the brick representation and the domain structure setup in \fig{fig:twoDomains}, the qualitative effects of refraction and reflection, which are the result of the anisotropies of the lattice, will be present irrespective of the representation. In the hexagonal representation where the anisotropies are less severe than in the brick representation, refraction and reflection only become significant at high energies. Similar reflective effects are also predicted with topological defects in cosmological field theories~\cite{vilenkin2000cosmic}. For example the conical nature of space around straight sections of cosmic strings can give rise to double images of galaxies or quasars~\cite{everett1974observational}. If such observations are ever to be found, domain boundaries may offer an alternative explanation. 

The observation of double imaging of astronomical objects implies domain sizes at commensurate scales, however it is not clear from our model what the average sizes of the domains should be. Although there is no precise method to predict the probable domain size in our model at present, it is likely related to the effective rate of cooling, as with conventional crystal properties. As fluctuations seed domains of random orientation in many different places, the rate of cooling will influence the ability of these smaller domains to coalesce into larger ones: Specifically a slower cooling rate would form larger domains than a faster rate. Many small domains with random orientations will form an amorphous manifold. With no preferred direction, on a scale much larger than the individual domains, this granular space will appear to exhibit rotational symmetry. On the other hand, at wavelength scales commensurate with the domain sizes the bosons will be scattered. In such a universe, double imaging of astronomical objects as a result of domain boundaries would not be observed. As our universe does appear to be isotropic and we do not see such scattering effects, the length scales of the domains must either be very small (sub-microscopic) or very large (astronomical).

\subsection{Domain lensing}
\label{sec:domain lensing}
\begin{figure}
	\includegraphics[width=\columnwidth]{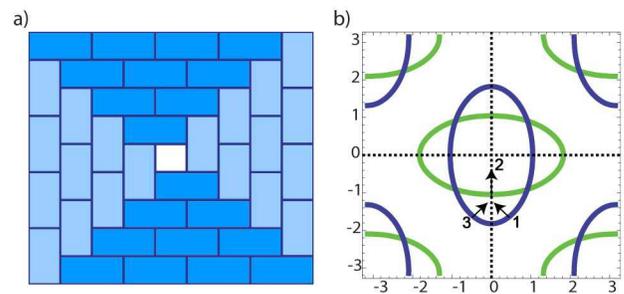}%
\caption{(Color online) a) The intersection of four domains (bottom, left, top, right) with orientation $\alpha=0,\pi/2$. b) Isoenergy contours at $E/\kappa=2$. The blue contour corresponds to domain-$0$ (bottom and top domains) and the green contour corresponds to domain-$\pi/2$ (left and right domains). The arrows represent group velocities. A mode originating in the bottom domain with $\mathbf{k}=(\pi/6,-\pi/2)$ propagates with $v_g^+/\kappa=(\sqrt{-3}/2, \sqrt{3}/2)$ (arrow 1). It couples to the propagating mode $\mathbf{k}=(0,-\pi/3)$ in the left domain (arrow 2). At the second domain boundary it will couple to the propagating mode $\mathbf{k}=(-\pi/2,-\pi/6)$ in the top domain (arrow 3). This trajectory will converge with its symmetric (about the $y$-axis) mode producing a lensing-like effect.}
\label{fig:four_domains}%
\end{figure}

Gravitational lensing is the bending of light as the result of gravitational effects such that multiple or distorted images are formed of the source. Light is bent most near the center of the lens as the gravitational strength here is greatest. Because of this, gravitational lenses are characterized by a focal line (as opposed to a focal point of conventional lenses). Here we show how a granular structure of space can give rise to similar lensing-like effects. This mechanism is distinct from the curved geodesics of general relativity. As an example we show that lensing-like effects can arise from the intersection of four (bottom, left, top, right) domains as illustrated in \fig{fig:four_domains}a). We assume the structure to extend indefinitely. Because every vertex has valence $v_0=3$, the domain structure of \fig{fig:four_domains}a) is stable to small fluctuations ($h=1$), as local edge hopping will not produce a lower energy state.

To understand how this four domain geometry can produce a lensing-like effect, we plot the isoenergy contours of the system at a particular frequency in \fig{fig:four_domains}b). The blue isoenergy contour corresponds to domain-0 and the green to domain-$\pi/2$. The interface of the domains occur at angles $\pm\pi/4$, requiring the phase matching condition $k_x - k_y = k_x' - k_y'$ at the bottom-left domain interface and $k_x' + k_y' = k_x'' + k_y''$ at left-top domain interface with similar conditions at the right domain interfaces. For example a mode originating in the bottom domain with $\mathbf{k}=(-\pi/6,-\pi/2)$ propagates with $\mathbf{v}_g^+/\kappa=(\sqrt{-3}/2, \sqrt{3}/2)$ (arrow 1). Phase matching at the domain boundary which is at $\pi/4$ relative to the $x$-axis, will mean it will couple to the propagating mode $\mathbf{k}=(0,-\pi/3)$ (arrow 2). At the second domain boundary which is at $-\pi/4$ relative to the $x$-axis, phase matching will mean that this mode will couple to the propagating mode $\mathbf{k}=(-\pi/2,-\pi/6)$ (arrow 3). This example is symmetric about the $x,y$-axis. So together with its symmetric mode reflected about the $x$-axis these trajectories will converge to focus the source. 

\begin{figure*}%
	\includegraphics[width=2\columnwidth]{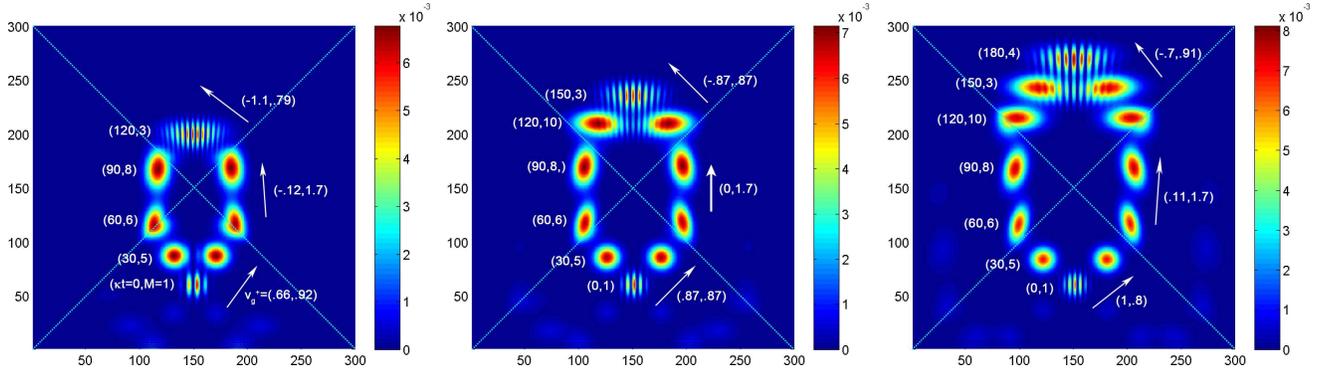}%
	\caption{(Color online) Simulation of boson Gaussian wavepackets with a) $\mathbf{k}=(\pm\pi/8,-\pi/8)$, b) $\mathbf{k}=(\pm\pi/6,-\pi/6)$, c)$\mathbf{k}=(\pm\pi/5,-\pi/5)$ at the four domain structure of \fig{fig:four_domains}. The arrows indicate the group velocity ($v_g^+$) in each domain. As the boson crosses the domain boundaries it undergoes refraction to converge on the far side. The foci points for the different modes of the boson wavepacket form a focal line. Note that there are also modes (the faint light blue pulses) associated with $v_g^-$ which propagate in the opposite direction (i.e. downwards) and are reflected off the hard wall boundary. The notation $(\kappa t, M)$ follows \fig{fig:refraction_simulation}.}%
\label{fig:lensing_simulation}%
\end{figure*}

The domain lensing effect is simulated by propagating a boson wavepacket in the lattice represented by \fig{fig:four_domains}a). In \fig{fig:lensing_simulation}a), the boson is initialized as a Gaussian pulse of two modes, $\mathbf{k}=(\pm\pi/8,-\pi/8)$, in the bottom domain. As it propagates through the domain boundaries (indicated by the dotted lines) it undergoes refraction, following the predicted trajectories, to focus on the far side. The simulation of other modes [\fig{fig:lensing_simulation}b),c)] show that their foci do not converge to the same point, but instead form a focal line. It is important to point out that this behavior, where a point source is separated and then focused again on a contiguous line (as determined by lattice connectivity) is a property independent of representation. In other words, this is a property of the interconnectivity of the domain structure of \fig{fig:four_domains}a) and not it's brick representation. How we presented this in \fig{fig:lensing_simulation} is of course a result of our choice of representation.

The presence of a focal line means that an observer in this 2D universe would detect a distorted image of the source.This effect can be compared to the distortion of images of astronomical objects due to gravitational lensing in our three dimensional world. It is important to note however that the effect in our model is purely non-relativistic. Furthermore, in GR gravitational lensing, light of all modes bends more the closer it is to the gravitational source; whereas in domain lensing (as illustrated in the preceding example) for a given mode the amount of bending (refraction angle) is invariant with the distance from the optical axis. This distinction offers a possible avenue in which to test the model.

\section{Conclusion}
Quantum graphity is a background independent model that provides an alternative viewpoint on the notion and structure of space, based on condensed matter concepts but extended to a dynamic quantum lattice.  Through an annealing process we explored metastable domain structures and boundary defects.  We investigated the propagation of bosons in these structures, revealing that they produce intriguing scattering, double imaging, and gravitational lensing-like effects.  Importantly this serves as a framework in which observable consequences of the QG model may allow it to be tested.

\section{Acknowledgments}
The authors would like to thank Ray R. Volkas for his feedback on the manuscript. J.Q.Q. and A.D.G. would like to thank Andrew L.C. Hayward on fundamental conceptual discussions. J.Q.Q. would like to thank Susan M. Heywood for support and general discussion. This work was supported by the Australian Research Council (ARC) under the Centre of Excellence scheme.  A.D.G. acknowledges the financial support of the ARC under Project DP0880466.

\section{Appendix}
Forming primitive unit cells as shown in \fig{fig:honeycomb_brick}b) means that each unit cell will have four nearest neighbors. We label the displacement between a cell and its four neighboring cells as $\mathbf{d}_{1}=(-d_x,d_y),\mathbf{d}_2=(d_x,d_y),\mathbf{d}_3=(d_x,-d_y),\mathbf{d}_4=(d_x,-d_y)$.The Hamiltonian in \eq{eq:submatrix_eq_2} is written as
\begin{equation}
\begin{split}
\sum_r&[H_{rs}]e^{i\mathbf{k}\cdot(\mathbf{D}_r-\mathbf{D}_s)} =\\
	&\kappa\begin{bmatrix}
	 0& 1+e^{i\mathbf{k}\cdot\mathbf{d}_2}+e^{i\mathbf{k}\cdot\mathbf{d}_4}\\ 
	 1+e^{i\mathbf{k}\cdot\mathbf{d}_1}+e^{i\mathbf{k}\cdot\mathbf{d}_3}& 0
	\end{bmatrix}~.
\end{split}
\label{eq:A1}
\end{equation}
Solving for the eigenvalues of this matrix, the dispersion relation is
\begin{equation}
	[E(\mathbf{k})/\kappa]^2 = 1+4\cos(k_xd_y)[\cos(k_xd_y)+\cos(k_yd_x)]~.
\label{eq:A2}
\end{equation}
In the brick representation $(d_x,d_y)=(1,1)$ and in the honeycomb representation $(d_x,d_y)=(3/2,\sqrt{3}/2)$.

\bibliography{bibliography}

\end{document}